# A Machine Learning Model for Early Detection of Diabetic Foot using Thermogram Images

Amith Khandakar[1, 2], Muhammad E. H. Chowdhury[1*], Mamun Bin Ibne Reaz[2*], Sawal Hamid Md Ali[2], Md Anwarul Hasan[3], Serkan Kiranyaz[1], Tawsifur Rahman[1], Rashad Alfkey[4], Ahmad Ashrif A. Bakar[2], Rayaz A. Malik[5]

[1]Department of Electrical Engineering, Qatar University, Doha-2713, Qatar

[2]Dept. of Electrical, Electronics and Systems Engineering, Universiti Kebangsaan Malaysia, Bangi, Selangor 43600, Malaysia

[3]Department of Industrial and Mechanical Engineering, Qatar University, Doha-2713, Qatar

[4]Acute Care Surgery and General Surgery, Hamad Medical Corporation, Qatar

[5]Weill Cornell Medicine-Qatar, Ar-Rayyan, Qatar

*Correspondence: Muhammad E.H. Chowdhury (mchowdhury@qu.edu.qa); Mamun Bin Ibne Reaz (mamun@ukm.edu.my)

**Abstract**

Diabetes foot ulceration (DFU) and amputation are a cause of significant morbidity. The prevention of DFU may be achieved by the identification of patients at risk of DFU and the institution of preventative measures through education and offloading. Several studies have reported that thermogram images may help to detect an increase in plantar temperature prior to DFU. However, the distribution of plantar temperature may be heterogeneous, making it difficult to quantify and utilize to predict outcomes. We have compared a machine learning-based scoring technique with feature selection and optimization techniques and learning classifiers to several state-of-the-art Convolutional Neural Networks (CNNs) on foot thermogram images and propose a robust solution to identify the diabetic foot. A comparatively shallow CNN model, MobilenetV2 achieved an F1 score of ~95% for a two-feet thermogram image-based classification and the AdaBoost Classifier used 10 features and achieved an F1 score of 97 %. A comparison of the inference time for the best-performing networks confirmed that the proposed algorithm can be deployed as a smartphone application to allow the user to monitor the progression of the DFU in a home setting.

# 1.Introduction

Diabetes Mellitus (DM) leads to major complications such as heart disease, stroke, renal failure, blindness, and diabetic foot ulceration (DFU) with lower limb amputation [1]. Healing of DFU can be difficult or delayed [2] with an increased risk of infection and amputation [3]. DFU recurs in approximately 40% of patients after the first year and in 60% after three years [4, 5] and leads to amputation in over 1 million diabetic patients annually in the USA [6]. In Europe, 250,000 diabetic patients undergo lower limb amputation with an associated mortality of 30% at one month and 50% at 1 year [7]. Diabetic foot ulceration is associated with markedly increased healthcare costs, decreased quality of life, infection, amputation, and death. The detection of patients at risk of DFU may enable timely intervention to prevent foot ulceration, amputation, and death.

Self-care via monitoring without medical assistance, for early signs of DFU, may allow timely offloading to prevent skin breakdown and development of a wound. Visual inspection has its limitations as people with obesity or visual impairment cannot see their site of ulceration. However, recent studies utilizing temperature monitoring have shown that they can predict the development of DFU in 97% of patients [4, 8-10]. Indeed, patients undergoing continuous foot temperature monitoring had a lower risk of DFU [11]. Skin temperature monitoring emerged during the 1970s, with "asymmetry analysis" proving to be very effective in identifying ulcers at an early stage [12]. A temperature difference of 2.22°C (4°F) over at least two consecutive days could be used as a threshold for therapy to prevent DFU [8]. The system correctly identified the development of DFU in 97% of participants, with an average lead time of 37 days [13].

Thermography is a rapid non-invasive imaging technique to quantify thermal changes in the diabetic foot [13]. Several studies have proposed thermogram-based techniques for identifying those at risk of DFU [2, 3, 14] by identifying a characteristic thermal distribution in the infrared image. The control group had a specific butterfly pattern [15] compared to a large variety of spatial patterns in the patients with diabetes [16, 17]. Whilst it is possible to assess thermal changes in one foot compared to the contralateral foot [18-21] if both feet have thermal changes without a butterfly pattern, then one foot cannot act as a reference. Asymmetry cannot be measured despite a large temperature difference and identical spatial distributions in both feet. An alternative approach is to calculate the temperature change with respect to the butterfly pattern of a control group [22-24].

Machine learning (ML) techniques have been widely used for automatic image classification using feature extraction, feature ranking, and using different ML models, such as Artificial Neural Networks (ANN), k-nearest neighbors (KNN), and Support Vector Machines (SVM) [25-27]. The change of focus from traditional paradigms in machine learning to Deep Learning (DL) is the product of the high accuracy achieved through its large learning structures, enabling DL to obtain deeper data traits. The need for large data size and high computational complexity can be addressed using transfer learning on pre-trained networks. Whilst it is reasonably straightforward to distinguish the foot thermogram of a control subject with a specific spatial pattern, the distribution in a diabetic foot without a specific spatial pattern is more challenging, especially as the spatial distribution may change and the detection of a temperature rise in the plantar region is important for diabetic patients.

Several studies [22, 23, 28-35] have attempted to extract features to identify the hot region in the plantar thermogram, to identify tissue damage or inflammation. Etehadtavakol et al. [35] proposed a method called lazy snapping to extract the extreme temperature areas in the thermogram images which can easily differentiate the coarse and fine-scale change. A thresholding method was used to identify the highest temperature areas from the plantar region [22], while Gururajarao et al. [34] used an active contour model of plantar segmentation and a thresholding method to extract the highest temperature points. Adam et al. [33] used Discrete Wavelet Transformation (DWT) and higher-order spectra (HOS) to derive several coefficients from the characteristics of texture and entropy. A double density-dual tree-complex wavelet transform (DD-DT-CWT) was used to decompose the image and extract several key features [32]. Saminathan et al. [31] segmented the plantar area into 11 regions using region-raising and extracted texture characteristics to classify it into a normal or ulcer group. Most of these works were reported on a small private dataset and utilized post-processing techniques, which might not be able to generialize on a different dataset and the real-time applicability and inference time were not reported. Moreover, the performance of these methods were not comparable to the machine learning based techniques.

Very few studies have applied the deep learning (DL) technique to classify thermogram images from controls and diabetic patients. Maldonado et al. [30] utilized the DL technique to segment the thermogram of the plantar area to classify ulceration or necrosis. Hernandez et al. [23] proposed a quantitative thermal change index (TCI) to measure the thermal change in the plantar region of diabetic patients to classify patients from controls. Hernandez et al. [23, 29] utilized the "Plantar Thermogram Database" of 334-foot thermogram images and used TCI to classify subjects into Class 1 to 5 based on the spatial temperature distribution and temperature range. Cruz-Vega et al. [28] also proposed a DL technique to classify the images of the 'Plantar Thermogram Database' into two classes at a time, but the technique is questionable as it cannot be used for clinical decision making and the applicability of such a solution for a smartphone application is not discussed.

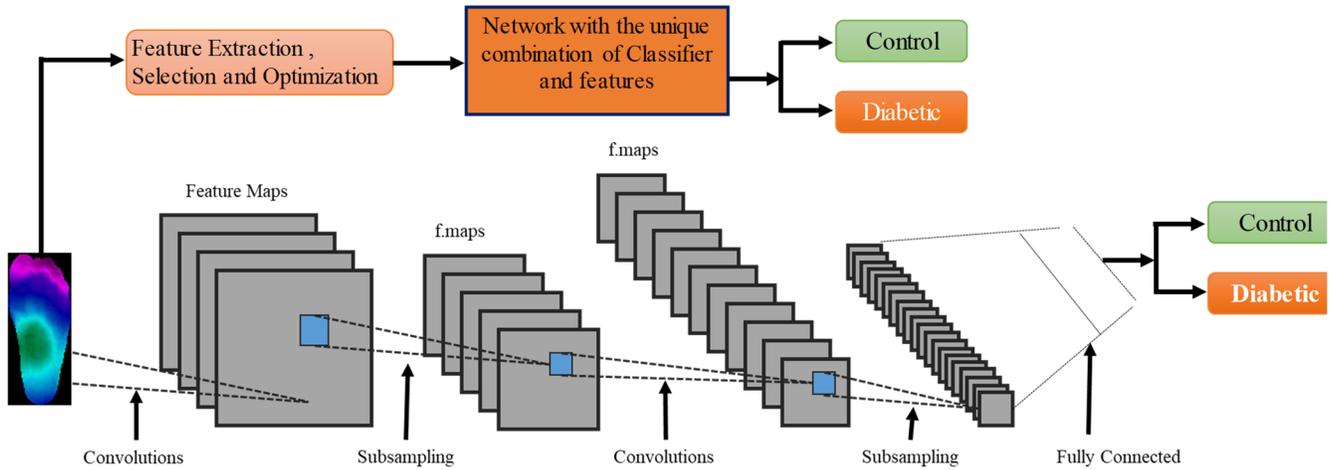

Figure 1. Proposed classification pipeline using one-dimensional (1D) machine learning classifiers and 2-dimensional (2D) thermogram image.

We have utilized an available dataset to classify control and diabetic groups and developed a novel technique to automatically classify the thermogram images and compared the outcome to a 2D deep learning technique. Moreover, the light architecture and machine learning model are deployable in smartphones.

The major contributions of this paper are:

- Comparative evaluation over the state-of-the-art 2D CNN models and image enhancement techniques for the detection of diabetic foot with high accuracy.
- A detailed investigation of the relevant features to improve the detection performance when used as input to traditional classifiers.
- An investigation of feature selection and optimization techniques and classification models to maximize detection performance utilizing light classifiers.

Section II discusses the methodology, section III presents the results and discussion and section IV presents the conclusions and proposes topics for future research.

## 2. Methodology

Figure 1 shows the complete system block diagram. The thermogram is used as an input to extract important features, feature optimization, and ranking by different ranking techniques. The best combination of the top-ranked features was used as input to the classifier to stratify the thermogram images into diabetic and control groups. The performance of the proposed technique was compared with a 2D CNN-based image classification model for comparative evaluation. Various image enhancement techniques were utilized to enhance the 2D thermogram images and improve the performance of 2D CNN [36].

DATASET DESCRIPTION

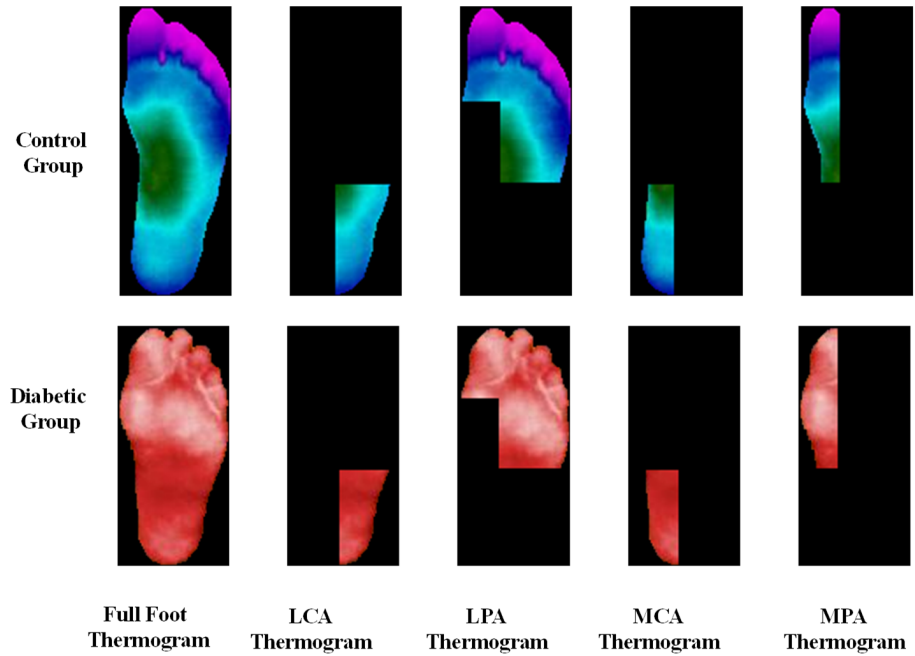

Figure 2: Sample of MPA, LPA, MCA, and LCA angiosomes for control and diabetic foot thermogram.

A database of age, gender, height, and weight and 167-foot pair thermograms from 122 participants with diabetes mellitus and 45 controls was made public by Hernandez-Contreras et al. [29]. Continuous variables were reported with the number of missing data, median, mean, and quartiles (Q1, Q3) for diabetic and control groups (Table I). The chi-square test was conducted for gender while the rank-sum test was conducted on other features. A p-value <0.05 was used as the cut-off for statistical signficance. The foot thermogram images were segmented to remove the background and were also segmented into four angiosomes for the medial plantar artery (MPA), lateral plantar artery (LPA), medial calcaneal artery (MCA), and lateral calcaneal artery (LCA) [37] (Figure 2). The angiosome related information is not only useful to identify the arteries associated with ulceration risk but also shows the local temperature of each angiosome. Pixelated temperature readings for the full foot and the four angiosomes for both feet were available in the dataset, to encounter the problem in two dimensions: pixelated temperature and the 2D thermogram image.

*2.1 Feature extraction from temperature map*

Different features have been extracted by different research groups from foot thermograms over the last decade. Cajacuri et al. [38] highlighted the importance of age, gender and body mass index. Contreras et al. [29] developed the thermal change index (TCI), the mean temperature difference between the corresponding angiosomes from a diabetic patient and a control group as shown in Equation (1).

$$TCI = \frac{CG_{ang} - DG_{ang}}{4} \tag{1}$$

where $CG_{ang}$ and $DG_{ang}$ are the temperature values of the angiosome for the control and diabetic groups, respectively. Barreto et al. [39] proposed features, such as Estimate temperature (ET), estimated temperature difference (ETD), and hot spot estimator (HSE) for analyzing thermograms, as shown in Equation (2)-(4).

$$ET = \frac{a_{j-1}C_{j-1} + a_j C_j + a_{j+1}C_{j+1}}{a_{j-1} + a_j + a_{j+1}} \tag{2}$$

$$ETD = |ET_{left\ Angiosome} - ET_{right\ angiosome}| \tag{3}$$

$$HSE = |C_l - ET| \tag{4}$$

To calculate these features, the temperature map in the thermogram image is categorized into temperature classes: $C_0$ to $C_7$. A histogram is generated for the percentage of pixels in the thermogram image, which lies in different temperature classes, where $C_0=26.5°C$, $C_1=28.5°C$, $C_2=29.5°C$, $C_3=30.5°C$, $C_4=31°C$, $C_5=32.5°C$, $C_6=33.5°C$, and $C_7=34.5°C$. The highest frequency of a temperature class is denoted by $C_j$ and the percentage of pixels in that region is $a_j$. The values $a_{j-1}$ and $a_{j+1}$ are the percentages of pixels in the neighboring temperature classes $C_{j-1}$ and $C_{j+1}$, respectively. The $a_j$ and $C_j$ values are used to calculate the ET of the thermogram for each angiosome which is used to calculate the ETD values. Finally, the HSE is calculated using ET and $C_l$ values, where $C_l$ is the highest temperature present in the angiosome regardless of its percentage in the histogram. HSE can identify severe DFU. Saminathan et al. [31] have stressed the importance of standard statistical parameters such as mean, standard deviation, and median used in various biomedical applications [40-42].

In addition to the above-mentioned features, features which are visually important to distinguish the variation in the plantar temperature distribution were formulated. Five distinct temperature ranges were found in the dataset and verified with the TCI parameters [29].

TABLE I. STATISTICAL ANALYSIS OF 1D FEATURES FOR BINARY CLASSIFICATION.

| | Item | Control | Diabetic | Total | Method | Statistic | P value |
|---|---|---|---|---|---|---|---|
| 1 | Gender | | | | Chi-square test | 39.3886 | P <.0001 |
| | • Male (%) | 58(64%) | 66(28%) | 124(37%) | | | |
| | • Female (%) | 32(36%) | 178(72%) | 210(63%) | | | |
| 2 | Age (Years) | | | | Rank-sum test | 12.6108 | P <.0001 |
| | • N (missing) | 90(0) | 244(0) | 334(0) | | | |
| | • Mean ± SD | 28±8 | 55.98±10.6 | 48.4±16 | | | |
| | • Median | 25 | 55 | 52 | | | |
| | • Q1, Q3 | 23,30 | 50,63 | 34, 60 | | | |
| | • Min, Max | 21,52 | 23,84 | 21,84 | | | |
| 3 | Full-Foot Temperature (°C) | | | | Rank-sum test | 7.6913 | P <.0001 |
| | • N (missing) | 90(0) | 244(0) | 334(0) | | | |
| | | 26.7±1.6 | 29.7±2.9 | 28.9±2.9 | | | |

|   |   |   | 26.8 | 30 | 28.8 |   |   |   |
|---|---|---|------|----|----|---|---|---|
|   |   | • Mean ± SD | 25.9, 27.7 | 27.9, 32 | 26.6, 31.2 |   |   |   |
|   |   | • Median | 22, 29.6 | 20.4, 35.6 | 20.4, 35.6 |   |   |   |
|   |   | • Q1, Q3 |   |   |   |   |   |   |
|   |   | • Min, Max |   |   |   |   |   |   |
| 4 | LCA Temperature (ºC) |   |   |   |   | Rank-sum test | 7.3563 | P <.0001 |
|   |   | • N (missing) | 90(0) | 244(0) | 334(0) |   |   |   |
|   |   | • Mean ± SD | 26.6±1.5 | 29.3±2.7 | 28.5±2.7 |   |   |   |
|   |   | • Median | 26.5 | 29.4 | 28.4 |   |   |   |
|   |   | • Q1, Q3 | 25.9, 27.6 | 27.4, 31.2 | 26.5, 30.6 |   |   |   |
|   |   | • Min, Max | 22.8, 30.1 | 20.9, 35.9 | 20.9, 35.3 |   |   |   |
| 5 | LPA Temperature (ºC) |   |   |   |   | Rank-sum test | 7.8004 | P <.0001 |
|   |   | • N (missing) | 90(0) | 244(0) | 334(0) |   |   |   |
|   |   | • Mean ± SD | 26.4±1.8 | 29.9±3.2 | 28.9±3.3 |   |   |   |
|   |   | • Median | 26.3 | 30.3 | 28.9 |   |   |   |
|   |   | • Q1, Q3 | 25.4, 27.6 | 27.6, 32.4 | 26.3, 31.8 |   |   |   |
|   |   | • Min, Max | 21.4, 30 | 19.8, 35.9 | 19.8, 35.9 |   |   |   |
| 6 | MCA Temperature (ºC) |   |   |   |   | Rank-sum test | 7.2299 | P <.0001 |
|   |   | • N (missing) | 90(0) | 244(0) | 334(0) |   |   |   |
|   |   | • Mean ± SD | 27±1.5 | 29.5±2.6 | 28.8±2.6 |   |   |   |
|   |   | • Median | 27.2 | 29.6 | 28.8 |   |   |   |
|   |   | • Q1, Q3 | 26.1, 28 | 27.8, 31.4 | 27, 30.8 |   |   |   |
|   |   | • Min, Max | 23, 30.2 | 21.3, 35.1 | 21.3, 35.1 |   |   |   |
| 7 | MPA Temperature (ºC) |   |   |   |   | Rank-sum test | 7.8193 | P <.0001 |
|   |   | • N (missing) | 90(0) | 244(0) | 334(0) |   |   |   |
|   |   | • Mean ± SD | 26.7±1.9 | 30.1±3.1 | 29.2±3.2 |   |   |   |
|   |   | • Median | 26.7 | 30.6 | 29.2 |   |   |   |
|   |   | • Q1, Q3 | 25.7, 27.9 | 28, 32.3 | 26.7, 31.8 |   |   |   |
|   |   | • Min, Max | 21.3, 30.5 | 20.3, 36.1 | 20.3, 36.1 |   |   |   |
| 8 | TCI Temperature (ºC) |   |   |   |   | Rank-sum test | 10.6670 | P <.0001 |
|   |   | • N (missing) | 90(0) | 244(0) | 334(0) |   |   |   |
|   |   | • Mean ± SD | 14±12.7 | 29.7±2.9 | 25.5±9.9 |   |   |   |
|   |   | • Median | 13 | 30 | 28.7 |   |   |   |
|   |   | • Q1, Q3 | 1.2, 26.8 | 27.8, 31.8 | 25.8, 31.1 |   |   |   |
|   |   | • Min, Max | 0.12, 29.6 | 20.6, 35.5 | 0.12, 35.5 |   |   |   |
| 9 | Outcome (%) | 90(27%) | 244(73%) | 334 |   |   |   |   |

Five distinct temperature ranges were classified into normalized temperature ranges (NTR). We have computed the variable $NRT_{class\,j}$ which is the number of pixels in *class j* temperature range over the total number of non-zero pixels, where *class j* can be class 1 to 5. For the temperature ranges in the class, we have used the same temperature range as reported in [29].

39 features were extracted for the early detection of diabetic foot, which are age, gender, TCI, highest temperature value, NTR (Class 1-5), HSE, ET, ETD, mean, median, SD of temperature for the different angiosomes: LPA, LCA, MPA, MCA, and the full foot.

The final list of features was optimized to remove redundant features by finding the correlation between the different features. Features with more than 95 % correlation were removed, which improves the overall performance by reducing the number of redundant features, avoiding overfitting [40-43].

*2.2. Classification using thermogram features*

Five-fold cross-validation was used in this study, where each fold was divided into a 80 % training and 20 % testing set. 20% of the training data was spared as the validation set. To avoid the issue of an imbalanced training dataset and biased estimates [44], Synthetic Minority Oversampling Technique (SMOTE) [45] was used for training data augmentation.

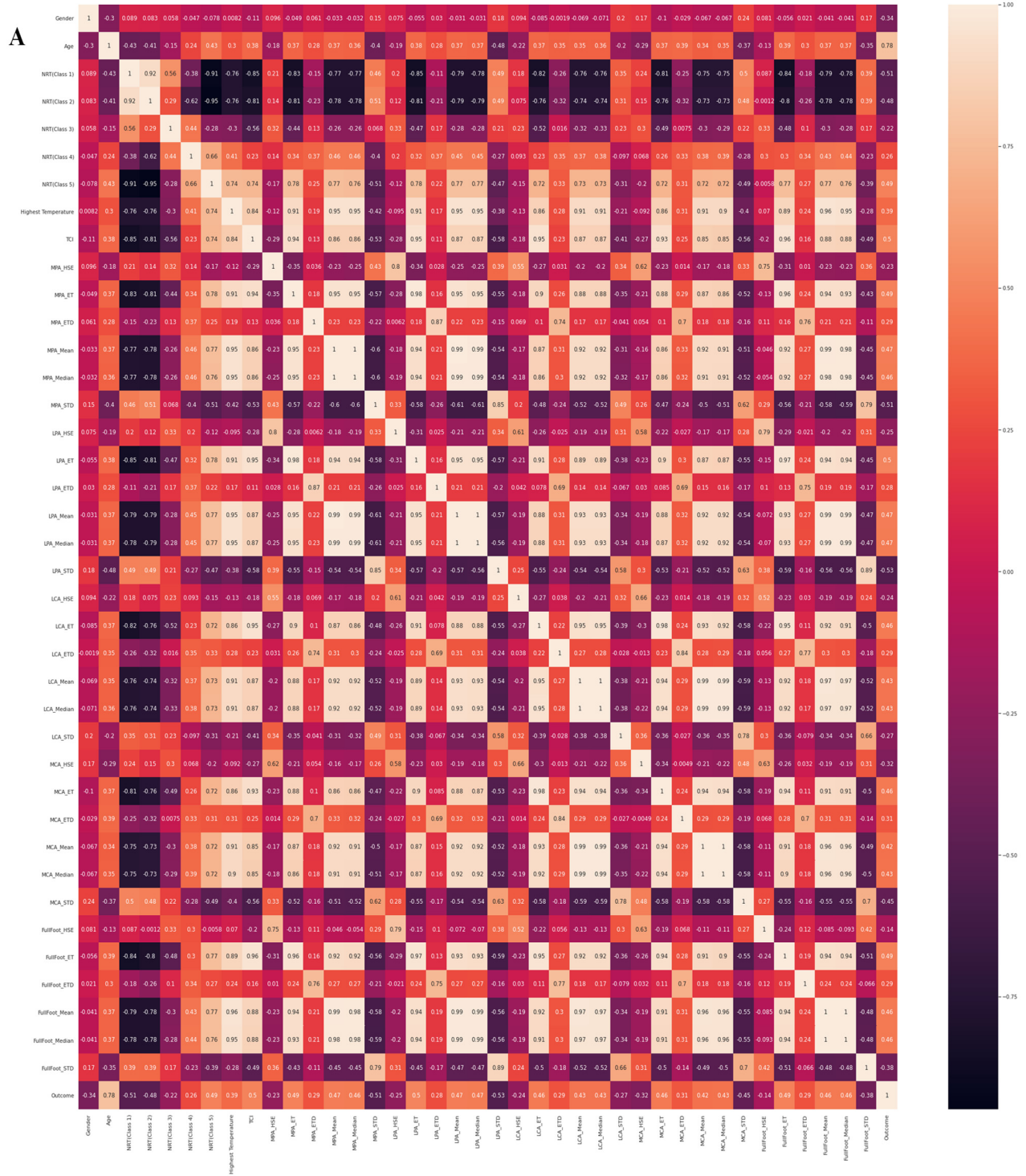

A

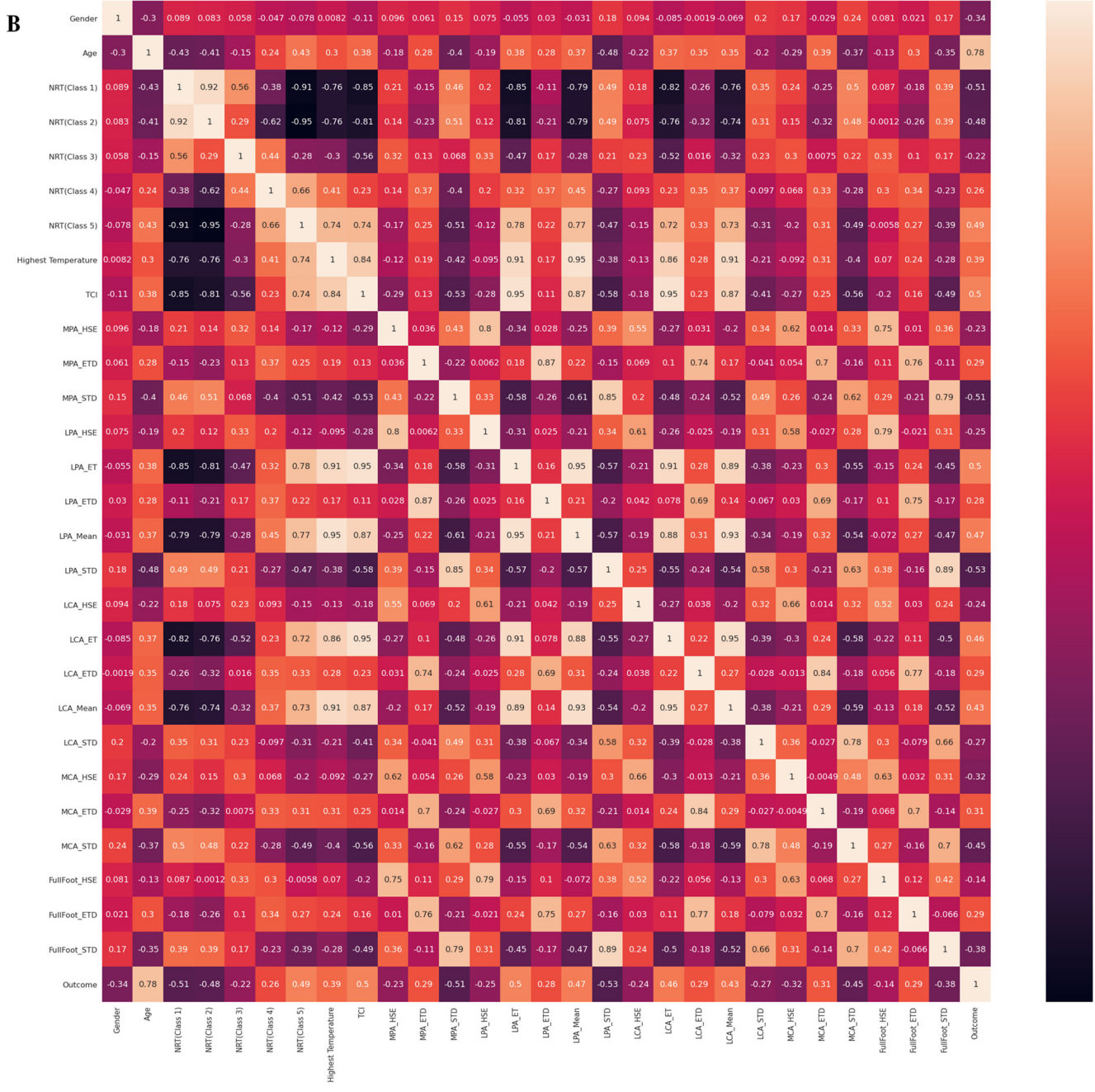

**Figure 3.** Heatmap of the correlation matrix with all the features (A) and after removing the highly correlated features (B).

*Feature Ranking Techniques:* The feature set was first optimized by removing any redundant features, i.e. features correlation more than 95% were removed. Of 39 features, after correlated feature reduction, the number of features became 28. The heatmap of the correlation matrix before and after removing the highly correlated features is presented in Figure 3. The reduced feature set used for further investigation was: age, gender, TCI, highest temperature value, NTR (Class 1-5), HSE, ETD, STD parameters for the different angiosomes, Full Foot, and ET, mean of LPA and LCA.

The shortlisted parameters of the dataset, after optimization, were assessed to take decisions and identify the top features for binary classification. Three different sets of feature ranking were identified using the Multi-Tree Extreme Gradient Boost (XGBoost) [46], Random Forest [47], and Extra Tree [48] techniques. Default parameters were used for the feature ranking techniques to avoid overfitting, a common problem with a large number of features and a limited sample size [49, 50]. The best performing top-ranked features from the different feature ranking techniques are used to identify the best combination of features using a rigorous investigation to identify the best combination of features that gave the best performance.

*Classifiers:* For a detailed investigation, different classifiers such as multilayer perceptron (MLP) [51], Logistic regression [52], K-Nearest Neighbor (KNN) [53], Adaboost [54], Support Vector Machine (SVM) [55], Random Forest [56], Extra Tree [57], Gradient Boosting [58], Extreme Gradient Boost (XGBoost) [59], Linear Discriminant Analysis (LDA) [60] were used. An MLP is characterized by several layers of neurons connected between the input and the output layers. MLP uses backpropagation for training the network. Logistic regression is a variant of regression function which uses a logistic function to model a binary dependent variable. While typical linear regression uses a linear relation between predictors and output, logistic regression uses a sigmoid function to relate output with linear prediction and linear prediction works like multivariate linear regression. KNN starts by determining "k", i.e., the number of neighbors to be compared. Once the paramter "k" is determined, the object's distance is computed with every object available in the dataset and the k-least distances were identified. XGBoost is the streamlined group calculation dependent on GBDT (Gradient Boosting Decision Tree). The principle concept of the boosting calculation is that numerous decision trees perform superior to a single one. LDA is a multi-class classification model, which can be used for dimensionality reduction. Random Forest is an ensemble of Decision Trees that combine the qualities of filter and wrapper methods. Extra Tree is a type of ensemble learning technique which aggregates the results of multiple de-correlated decision trees collected in a "forest" to output it's classification result. It is similar to a Random Forest and only differs from it in the manner of construction of the decision trees in the forest. AdaBoost classifier is a meta-estimator that begins by fitting a classifier on the original dataset and then fits additional copies of the classifier on the same dataset and adjusted focusing more on difficult cases.

In this experiment, 3 feature selection techniques with 10 machine learning models were investigated with 28 optimized features to identify the best-combined results in 840 investigations.

*2.3. Thermogram Image Classification by 2D CNNs*

The application of 2D CNNs in biomedical applications is popular for automatic and early detection of abnormalities such as COVID-19 pneumonia [61-63], Tuberculosis [64], community acquired pneumonia [65], and many others [66]. As before, five-fold cross-validation is applied, i.e. the dataset is divided into five-folds, and performance metrics were reported for cumulative folds. Overall accuracy and weighted average of Precision, Sensitivity, Specificity, and F1-Score are reported. Since the binary class dataset is not balanced and the number of images in 80% of the dataset (training set per fold) was small. The training dataset was augmented using image rotation and translation [61-65]. The details of the training, validation and testing dataset for 2D binary classification are presented in Table II.

Table II

Details of the dataset used for training, validation, and testing.

| Dataset | Class | Training Dataset Details | | | |
|---|---|---|---|---|---|
| | | Training Data/ Fold | Augmented Training Data/ Fold | Validation Data/ Fold | Test Data/ Fold |
| Contreras et al. [29] | DM | 190 | 1330 | 8 | 46 |
| | CG | 64 | 1664 | 4 | 22 |

*Transfer Learning:* Since the dataset size is small, pre-trained models, originally trained on the ImageNet database [67] were used in this study. Based on an extensive literature review and previous work [61-65], six well-known pre-trained deep learning CNNs were used in this study: ResNet18, ResNet50 [68], DenseNet201[68], InceptionV3 [69], VGG19 [70] and MobileNetV2 [71]

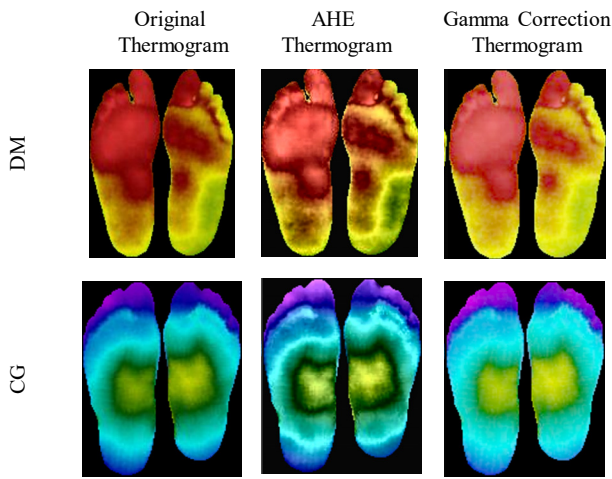

Figure 4. Original versus enhanced thermogram images using adaptive histogram equalization (AHE) and Gamma Correction for control and diabetic foot thermograms.

*Image Enhancement:* Image enhancement techniques such as Histogram Equalization (HE) [72], and Adaptive Histogram Equalization (AHE), and Gamma correction [73] can help 2D CNN in classification performances [36]. We have used the AHE technique (Figure 3) which performs histogram equalization over small regions (i.e., patches) in the image to enhance the contrast of each region individually. It improves local contrast and edges adaptively in each region of the image to the local distribution of pixel intensities instead of the global information of the image. Gamma correction was also applied to enhance the thermogram images. It performs a non-linear operation on the source image pixels, which alternates the pixel value to improve the image using the projection relationship between the value of the pixel and the value of the gamma according to the internal map. Sample thermogram image for DM and CG patients and the enhanced images with AHE and gamma correction are shown in Figure 4.

*2.4 Performance metrics*

Six performance metrics: Sensitivity, Specificity, Precision, Accuracy, F1-Score, and Area under the curve (AUC) were used as evaluation metrics where TP, FP, TN and FN are True Positive, False Positive, True Negative, and False Negative, respectively.

$$Sensitivity = \frac{(TP)}{(TP+FN)} \tag{5}$$

$$Specificity = \frac{(TN)}{(TN+FP)} \tag{6}$$

$$Precision = \frac{(TP)}{(TP+FP)} \tag{7}$$

$$Accuracy = \frac{TP+TN}{(TP+FN)+(FP+TN)} \tag{8}$$

$$F1\ Score = \frac{(2*Precision*Sensitivity)}{(Precision+Sensitivity)} \tag{9}$$

TP is the number of thermograms correctly identified as DM, FP is the number of incorrectly identified thermograms as DM, TN is the number of thermograms correctly classified as CG, and FN is the number of thermograms incorrectly identified as CG. We report the overall accuracy and weighted performance metric, with a 95 % confidence interval (CI), for Sensitivity, Specificity, Precision, and F1 Score. In addition, to compare the computational complexity of the different machine learning techniques, the inference time was calculated for the best performing 2D CNN models and 1D classifiers. The models that can be deployed in a smartphone were also identified.

All the experiments were performed by a computer with the following configuration: CPU Intel i7–10750H @2.6 GHz, GPU NVIDIA GeForce RTX 2070 Super, RAM 32 GB. Matlab 2020a was used for initial pre-processing and scikit-learn and PyTorch were used for classical machine learning and deep learning models, respectively.

## 3. Results and Discussion

The experimental results are divided into two sections: The first section presents the foot ulcer detection results by deep CNN models with transfer learning over the pre-trained networks while exploring the effects of different image enhancement techniques on thermogram image classification. Moreover, the effect of single and dual-foot as input was investigated for binary classification. In the second section, comparative evaluations among the best-performing machine learning models on the optimized thermogram features are presented.

Table III

Performance metrics for the binary classification using a single foot thermogram using 2D CNN. The best-performing network is highlighted in bold.

| Network | Class | Accuracy (%) | Precision (%) | Sensitivity (%) | F1-score (%) | Specificity (%) | Inference time (msec) |
|---|---|---|---|---|---|---|---|
| MobilenetV2 | DM | 92.51 ± 5.44 | 94.69 ± 4.63 | 95.08 ± 4.47 | 94.88 ± 4.55 | 85.56 ± 7.26 | |
|  | CG | 92.51 ± 3.30 | 86.52 ± 4.29 | 85.56 ± 4.41 | 86.04 ± 4.35 | 95.08 ± 2.71 | 5.252 |
|  | Overall | 92.51 ± 2.82 | 92.49 ± 2.83 | 92.51 ± 2.82 | 92.50 ± 2.82 | 88.13 ± 3.47 | |
| Resnet18 | DM | 90.42 ± 6.08 | 91.41 ± 5.79 | 95.90 ± 4.10 | 93.60 ± 5.06 | 75.56 ± 8.88 | |
|  | CG | 90.42 ± 3.69 | 87.18 ± 4.19 | 75.56 ± 5.39 | 80.96 ± 4.93 | 95.90 ± 2.49 | 2.545 |
|  | Overall | 90.42 ± 3.16 | 90.27 ± 3.18 | 90.42 ± 3.16 | 90.19 ± 3.19 | 81.04 ± 4.20 | |
| Resnet50 | DM | 93.41 ± 5.13 | 94.05 ± 4.89 | 97.13 ± 3.45 | 95.57 ± 4.25 | 83.33 ± 7.70 | |
|  | CG | 93.41 ± 3.11 | 91.46 ± 3.51 | 83.33 ± 4.68 | 87.21 ± 4.19 | 97.13 ± 2.09 | 6.164 |
|  | Overall | 93.41 ± 2.66 | 93.35 ± 2.67 | 93.41 ± 2.66 | 93.32 ± 2.68 | 87.05 ± 3.60 | |
| **DenseNet201** | **DM** | **94.01 ± 4.91** | **95.91 ± 4.11** | **95.91 ± 4.11** | **95.91 ± 4.11** | **88.89 ± 6.49** | |
|  | **CG** | **94.01 ± 2.98** | **88.89 ± 3.94** | **88.89 ± 3.94** | **88.89 ± 3.94** | **95.91 ± 2.49** | **26.138** |
|  | **Overall** | **94.01 ± 2.54** | **94.01 ± 2.54** | **94.01 ± 2.54** | **94.01 ± 2.54** | **90.78 ± 3.11** | |
| InceptionV3 | DM | 93.71 ± 5.02 | 93.73 ± 5.01 | 97.95 ± 2.93 | 95.79 ± 4.15 | 82.22 ± 7.90 | |
|  | CG | 93.71 ± 3.05 | 93.67 ± 3.06 | 82.22 ± 4.80 | 87.57 ± 4.14 | 97.95 ± 1.78 | 15.353 |
|  | Overall | 93.71 ± 2.60 | 93.71 ± 2.60 | 93.71 ± 2.60 | 93.58 ± 2.63 | 86.46 ± 3.67 | |
| VGG19 | DM | 92.22 ± 5.53 | 93.60 ± 5.06 | 95.90 ± 4.10 | 94.74 ± 4.61 | 82.22 ± 7.90 | |
|  | CG | 92.22 ± 3.36 | 88.10 ± 4.06 | 82.22 ± 4.80 | 85.06 ± 4.47 | 95.90 ± 2.49 | 6.284 |
|  | Overall | 92.22 ± 2.87 | 92.12 ± 2.89 | 92.21 ± 2.87 | 92.13 ± 2.89 | 85.91 ± 3.73 | |

*3.1 Detection results by deep CNN models*

The detection results of six deep CNN models for classifying the thermograms into control and diabetic groups from a single foot thermogram without and with image enhancement are presented in Table III and IV; while the data for both feet are shown in Table V and VI, respectively. It can be seen that the original thermograms perform better than the image enhancement techniques (AHE and Gamma), (Table IV) using the single foot thermogram. Among the six different deep CNN models investigated, DenseNet201 outperforms other networks with overall 94.01% sensitivity for the detection of DF and the class-wise sensitivities are 95.9% and 88.89% for DM and CG, respectively.

Table IV

Performance metrics for the best performing networks using 2D CNN on different image enhancement techniques using single foot thermograms. The best-performing network is highlighted in bold.

| Enhancement Technique | Best Network | Class | Accuracy (%) | Precision (%) | Sensitivity (%) | F1-score (%) | Specificity (%) | Inference time (msec) |
|---|---|---|---|---|---|---|---|---|
| **Original** | **DenseNet201** | **DM** | **94.01 ± 4.91** | **95.91 ± 4.11** | **95.91 ± 4.11** | **95.91 ± 4.11** | **88.89 ± 6.49** | **26.138** |
| | | **CG** | **94.01 ± 2.98** | **88.89 ± 3.94** | **88.89 ± 3.94** | **88.89 ± 3.94** | **95.91 ± 2.49** | |
| | | **Overall** | **94.01 ± 2.54** | **94.01 ± 2.54** | **94.01 ± 2.54** | **94.01 ± 2.54** | **90.78 ± 3.11** | |
| AHE | InceptionV3 | DM | 92.22 ± 5.53 | 94.67 ± 4.64 | 94.67 ± 4.64 | 94.67 ± 4.64 | 85.56 ± 7.26 | 15.450 |
| | | CG | 92.22 ± 3.36 | 85.56 ± 4.41 | 85.56 ± 4.41 | 85.56 ± 4.41 | 94.67 ± 2.82 | |
| | | Overall | 92.22 ± 2.87 | 92.22 ± 2.87 | 92.22 ± 2.87 | 92.22 ± 2.87 | 88.01 ± 3.48 | |
| Gamma Correction | InceptionV3 | DM | 93.41± 6.44 | 93.70± 6.30 | 97.54± 6.30 | 95.58± 6.30 | 82.22± 9.92 | 15.422 |
| | | CG | 93.41± 6.12 | 92.51± 5.30 | 82.22± 5.30 | 87.06± 5.30 | 97.54± 3.30 | |
| | | Overall | 93.41± 3.11 | 93.38± 3.12 | 93.41± 3.12 | 93.28± 3.12 | 86.35± 4.30 | |

Table V

Performance metrics for the binary classification using Gamma enhanced dual-foot thermogram using deep CNNs. The best-performing network is highlighted in bold.

| Network | Class | Accuracy (%) | Precision (%) | Sensitivity (%) | F1-score (%) | Specificity (%) | Inference time (msec) |
|---|---|---|---|---|---|---|---|
| **MobilenetV2** | **DM** | **95.81 ± 4.14** | **97.52 ± 3.21** | **96.72 ± 3.68** | **97.12 ± 3.46** | **93.33 ± 5.15** | **5.188** |
| | **CG** | **95.81 ± 2.51** | **91.30 ± 3.54** | **93.33 ± 3.13** | **92.30 ± 3.35** | **96.72 ± 2.23** | |
| | **Overall** | **95.81 ± 2.15** | **95.84 ± 2.14** | **95.81 ± 2.15** | **95.82 ± 2.15** | **94.24 ± 2.50** | |
| Resnet18 | DM | 93.41 ± 5.13 | 94.40 ± 4.75 | 96.72 ± 3.68 | 95.55 ± 4.26 | 84.44 ± 7.49 | 2.430 |
| | CG | 93.41 ± 3.11 | 90.48 ± 3.68 | 84.44 ± 4.55 | 87.36 ± 4.17 | 96.72 ± 2.23 | |
| | Overall | 93.41 ± 2.66 | 93.34 ± 2.67 | 93.41 ± 2.66 | 93.34 ± 2.67 | 87.75 ± 3.52 | |
| Resnet50 | DM | 90.42 ± 6.08 | 92.74 ± 5.36 | 94.26 ± 4.81 | 93.49 ± 5.1 | 80.00 ± 8.26 | 6.164 |
| | CG | 90.42 ± 3.69 | 83.72 ± 4.63 | 80.00 ± 5.02 | 81.82 ± 4.84 | 94.26 ± 2.92 | |
| | Overall | 90.42 ± 3.16 | 90.31 ± 3.17 | 90.42 ± 3.16 | 90.35 ± 3.17 | 83.84 ± 3.95 | |
| DenseNet201 | DM | 91.62 ± 5.72 | 92.86 ± 5.32 | 95.90 ± 4.10 | 94.36 ± 4.77 | 80.00 ± 8.26 | 25.732 |
| | CG | 91.62 ± 3.48 | 87.80 ± 4.11 | 80.00 ± 5.02 | 83.72 ± 4.63 | 95.90 ± 2.49 | |
| | Overall | 91.62 ± 2.97 | 91.50 ± 2.99 | 91.62 ± 2.97 | 91.49 ± 2.99 | 84.28 ± 3.90 | |
| InceptionV3 | DM | 93.41 ± 5.13 | 93.70 ± 5.02 | 97.54 ± 3.20 | 95.58 ± 4.25 | 82.22 ± 7.90 | 16.701 |
| | CG | 93.41 ± 3.11 | 92.50 ± 3.30 | 82.22 ± 4.80 | 87.06 ± 4.21 | 97.54 ± 1.94 | |
| | Overall | 93.41 ± 2.66 | 93.38 ± 2.67 | 93.41 ± 2.66 | 93.28 ± 2.69 | 86.35 ± 3.68 | |
| VGG19 | DM | 92.22 ± 5.53 | 92.91 ± 5.30 | 96.72 ± 3.68 | 94.78 ± 4.60 | 80.00 ± 8.26 | 6.292 |
| | CG | 92.22 ± 3.36 | 90.00 ± 3.76 | 80.00 ± 5.02 | 84.71 ± 4.52 | 96.72 ± 2.23 | |
| | Overall | 92.22 ± 2.87 | 92.13 ± 2.89 | 92.21 ± 2.87 | 92.07 ± 2.90 | 84.51 ± 3.88 | |

Table VI

Performance metrics for the best-performing networks using 2D CNN on different image enhancement techniques of combined foot thermograms.

| Enhancement Technique | Network | Class | Accuracy | Precision | Sensitivity | F1-score | Specificity | Inference time (msec) |
|---|---|---|---|---|---|---|---|---|
| Original | DenseNet201 | DM | 90.72 ± 5.99 | 93.47 ± 5.10 | 93.85 ± 4.96 | 93.66 ± 5.03 | 82.22 ± 7.90 | 24.362 |
| | | CG | 90.72 ± 3.64 | 83.15 ± 4.70 | 82.22 ± 4.80 | 82.68 ± 4.75 | 93.85 ± 3.01 | |
| | | Overall | 90.72 ± 3.11 | 90.69 ± 3.12 | 90.72 ± 3.11 | 90.70 ± 3.11 | 85.35 ± 3.79 | |
| AHE | MobilenetV2 | DM | 92.22 ± 5.53 | 94.67 ± 4.64 | 94.67 ± 4.64 | 94.67 ± 4.64 | 85.56 ± 7.26 | 5.363 |
| | | CG | 92.22 ± 3.36 | 85.56 ± 4.41 | 85.56 ± 4.41 | 85.56 ± 4.41 | 94.67 ± 2.82 | |
| | | Overall | 92.22 ± 2.87 | 92.22 ± 2.87 | 92.22 ± 2.87 | 92.22 ± 2.87 | 88.01 ± 3.48 | |
| Gamma Correction | MobilenetV2 | DM | 95.81 ± 5.20 | 97.52 ± 4.04 | 96.72 ± 4.62 | 97.12 ± 4.34 | 93.33 ± 6.48 | **5.188** |
| | | CG | 95.81 ± 4.95 | 91.30 ± 6.96 | 93.33 ± 6.16 | 92.30 ± 6.58 | 96.72 ± 4.40 | |
| | | Overall | **95.81 ± 2.51** | **95.84 ± 2.51** | **95.81 ± 2.51** | **95.82 ± 2.51** | **94.24 ± 2.92** | |

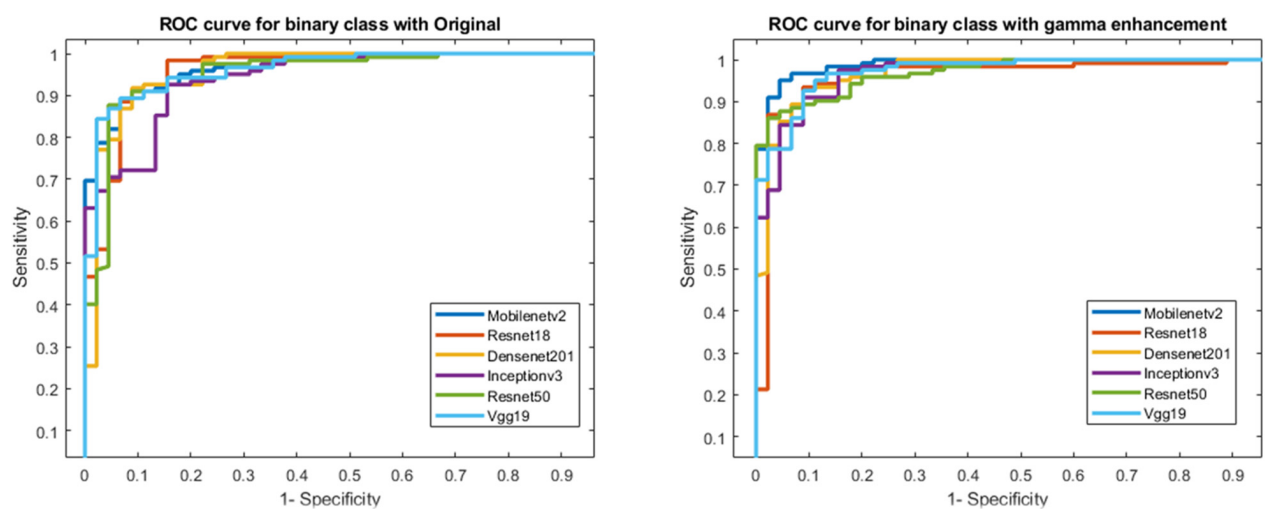

Figure 5. ROC for the Original and Gamma Correction Enhanced thermogram using Combined Foot Thermograms.

We have further investigated whether or not using a combination of foot images improves the detection performance. It was found that the Gamma enhanced dual-foot thermogram has outperformed the other methods (Table V). Interestingly, shallow network MobilenetV2 provides the best performance with an overall 95.81% sensitivity for diabetic foot detection and the class-wise sensitivities are 96.72% and 93.33% for DM and CG, respectively.

The outperformance using a combination of foot images is explained by the fact that combined foot thermograms provide more distinguishable features which are further enhanced by the image enhancement techniques.

Figure 5 clearly shows that the utilization of Gamma enhanced thermograms improved the classification performance compared to the original thermogram images for dual-foot investigation.

## 3.2 Feature-based detection results

We have investigated the performance of the 10 traditional classifiers with the three feature selection techniques and different combinations of optimized features. The summary of the top-performing five combinations is presented in Table VII. It can be seen that the AdaBoost Classifier with Random Forest Feature selection technique and the top 10 features shows the best performance of 96.71% sensitivity for diabetic foot detection and the class-wise sensitivities are 97.75% and 93.85% for DM and CG, respectively which is better than the top performance achieved by the deep CNN models.

TABLE VII

Performance metrics for the best-performing combinations.

| Classifier | Feature Selection | # of Feature | Class | Accuracy | Precision | Sensitivity | F1-score | Specificity | Inference time (ms) |
|---|---|---|---|---|---|---|---|---|---|
| **AdaBoost** | **Random Forest** | 10 | DM | 96.71 ± 3.69 | 97.55 ± 3.19 | 97.95 ± 2.93 | 97.75 ± 3.06 | 93.33 ± 5.15 | 0.397 |
|  |  |  | CG | 96.71 ± 2.24 | 94.38 ± 2.89 | 93.33 ± 3.13 | 93.85 ± 3.01 | 97.95 ± 1.78 |  |
|  |  |  | Overall | 96.71 ± 1.91 | 96.70 ± 1.92 | 96.71 ± 1.91 | 96.70 ± 1.92 | 94.58 ± 2.43 |  |
| AdaBoost | Extra Tree | 12 | DM | 96.41 ± 3.85 | 98.33 ± 2.64 | 96.72 ± 3.68 | 97.52 ± 3.21 | 95.56 ± 4.26 | 0.441 |
|  |  |  | CG | 96.41 ± 2.34 | 91.49 ± 3.50 | 95.56 ± 2.59 | 93.48 ± 3.10 | 96.72 ± 2.23 |  |
|  |  |  | Overall | 96.41 ± 2.00 | 96.49 ± 1.97 | 96.41 ± 2.00 | 96.43 ± 1.99 | 95.87 ± 2.13 |  |
| AdaBoost | Random Forest | 17 | DM | 96.41 ± 3.85 | 97.93 ± 2.94 | 97.13 ± 3.45 | 97.53 ± 3.21 | 94.44 ± 4.73 | 0.519 |
|  |  |  | CG | 96.41 ± 2.34 | 92.39 ± 3.33 | 94.44 ± 2.87 | 93.41 ± 3.11 | 97.13 ± 2.09 |  |
|  |  |  | Overall | 96.41 ± 2.00 | 96.44 ± 1.99 | 96.41 ± 2.00 | 96.42 ± 1.99 | 95.17 ± 2.30 |  |
| AdaBoost | Random Forest | 19 | DM | 96.41 ± 3.85 | 97.93 ± 2.94 | 97.13 ± 3.45 | 97.53 ± 3.21 | 94.44 ± 4.73 | 0.420 |
|  |  |  | CG | 96.41 ± 2.34 | 92.39 ± 3.33 | 94.44 ± 2.87 | 93.41 ± 3.11 | 97.13 ± 2.09 |  |
|  |  |  | Overall | 96.41 ± 2.00 | 96.44 ± 1.99 | 96.41 ± 2.00 | 96.42 ± 1.99 | 95.17 ± 2.3 |  |
| Extra Tree | Extra Tree | 8 | DM | 96.11 ± 4.00 | 97.93 ± 2.94 | 96.72 ± 3.68 | 97.32 ± 3.34 | 94.44 ± 4.73 | 0.299 |
|  |  |  | CG | 96.11 ± 2.43 | 91.40 ± 3.52 | 94.44 ± 2.87 | 92.90 ± 3.22 | 96.72 ± 2.23 |  |
|  |  |  | Overall | 96.11 ± 2.07 | 96.17 ± 2.06 | 96.11 ± 2.07 | 96.13 ± 2.07 | 95.06 ± 2.32 |  |

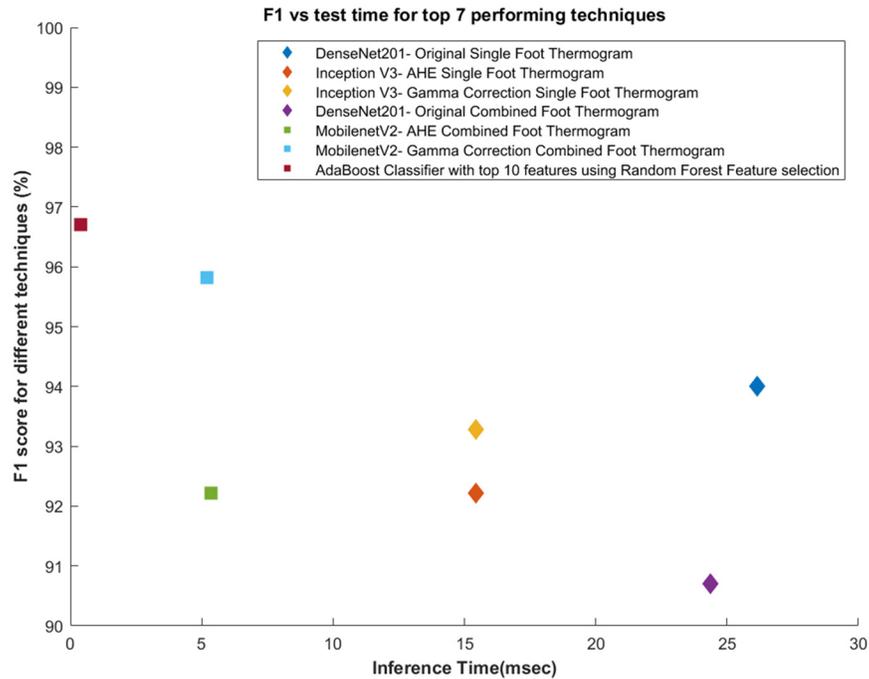

**Figure 6.** Comparison of F1-score versus Inference time for the top 7 performing techniques. Note:

The top-performing networks that can be deployed on smart portable devices are shown as Square blocks while Diamond blocks represent non-deployable models.

Figure 6 shows the comparison of the F1-score and inference time for the top 7 performing machine learning techniques from each category- i) different image enhancement on single foot thermogram, ii) different image enhancement on combined foot thermogram, and iii) the best performing 1D classifier, respectively. Only MobileNetv2 among the CNN models and AdaBoost classifier are deployable in the mobile platform.

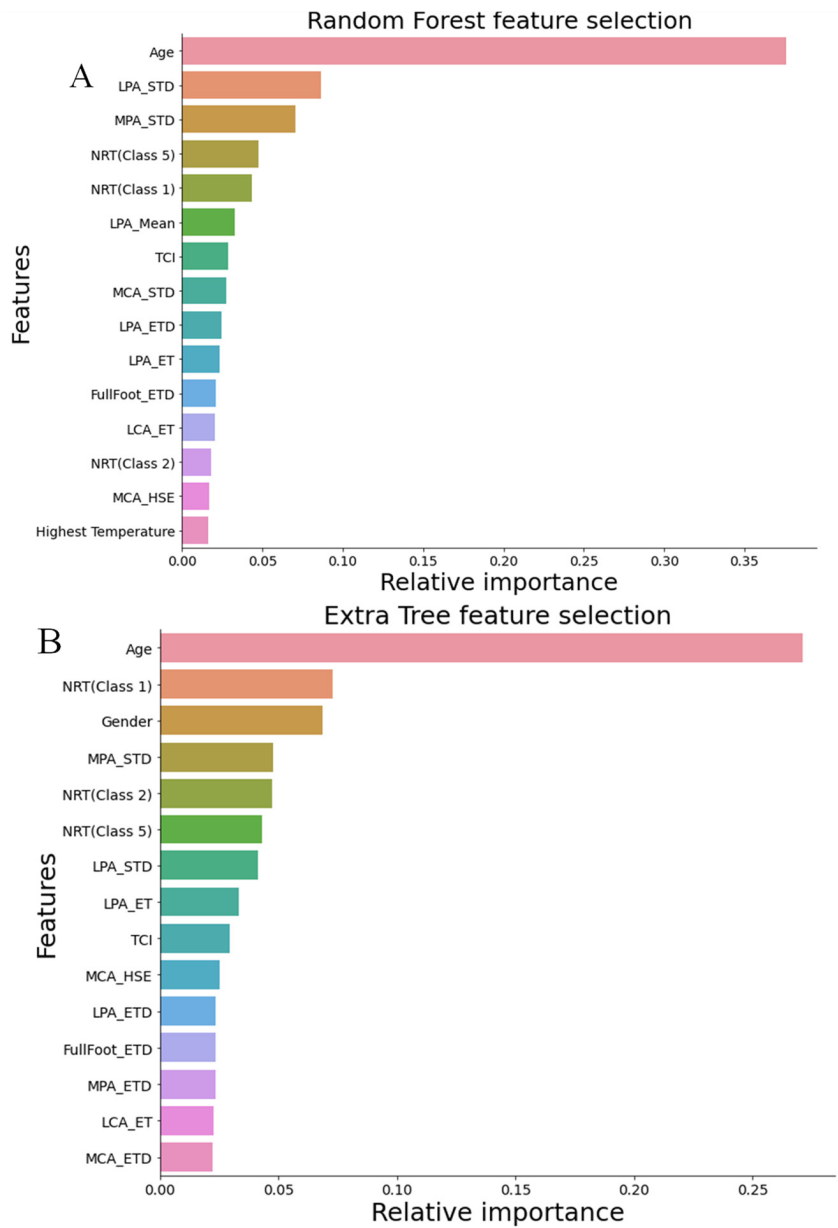

**Figure 7.** Top 15 features using A) Random Forest and B) Extra Tree feature selection techniques.

To the best of the author's knowledge, this is the first detailed investigation for diabetic foot detection using deep CNN models versus traditional machine learning approaches. All possible combinations in terms of classifier and feature selection techniques, along with the ranked features were investigated. As can be seen from Table VII, the Adaboost classifier outperforms other classifiers and the random forest feature ranking technique provides the best feature combination. The top 15 features among 28 features using Random Forest and Extra Tree feature selection techniques, after removing the highly correlated features from the initial 39 features, are shown in Figure 7. It is evident from Table VII that AdaBoost with the top 10 features (Age, LPA_STD, MPD_STD, NRT (Class 1), NRT (Class 5), LPA_mean, TCI, MCA_STD, LPA_ETD, and LPA_ET) has achieved the best classification performance.

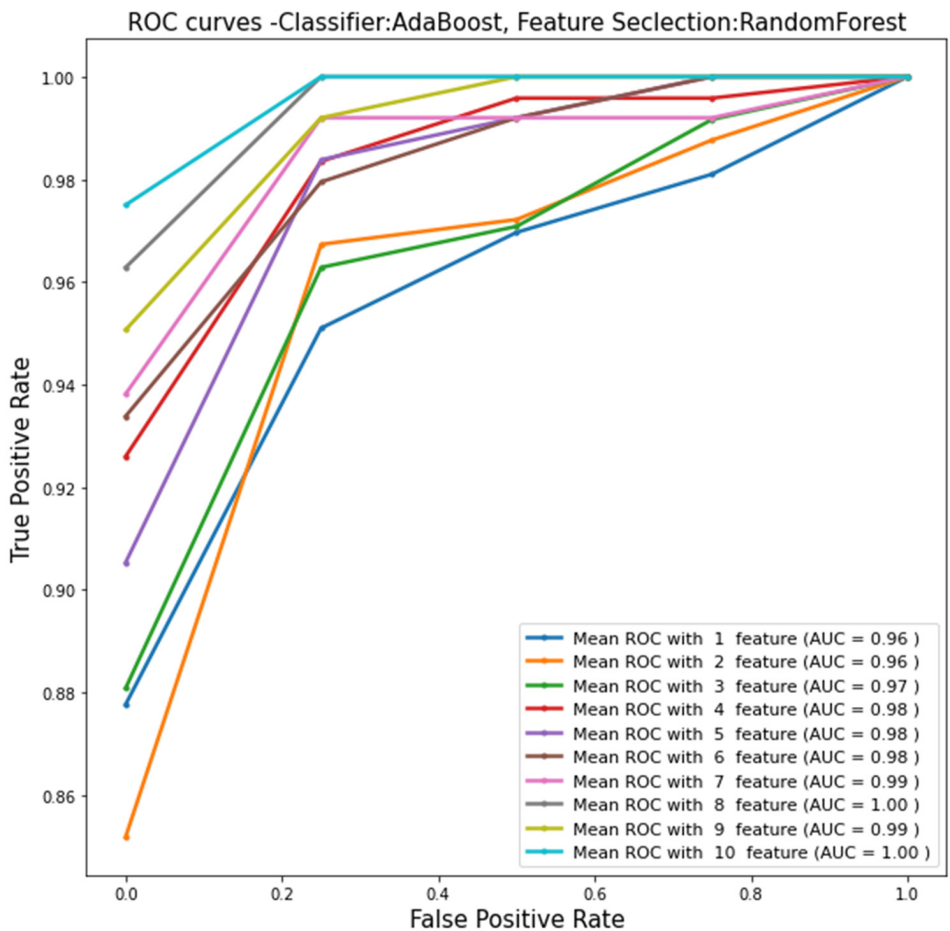

**Figure 8.** ROC curves for the top 10 feature combinations.

It should be noted that the feature-based classification was done using single foot thermogram features which outperform the dual-foot approach of enhanced image thermogram using deep CNNs. However, in the feature-based approach, demographic information such as age helps to improve its performance as reported in previous work [38]. Peregrine et al. [39] have identified 11 regions of interest (ROI), which can be used to identify the diabetic foot with the help of ET, ETD, and HSE. Figure 8 demonstrating the

ROC curves for the top 1 to 10 feature combinations also confirm that the top 10 feature combinations provided the best AUC.

As the hallux/big toe is a prominent region of interest and is in the LPA section of the foot, its contribution in the classification of the foot into diabetic and control is vital in the classification and it is natural to be included in the top 10 features. Age is a strong predictor of the diabetic foot as observed in this study [38]. Minor temperature variation is typically expected in the feet, but less variation, indicated by a lower standard deviation of temperature in LPA and MCA angiosomes, can also be an indicator of the diabetic foot. TCI is also an important indicator as it is a summary of the temperature variation in all angiosomes.

To the best of our knowledge, no previous study has reported an image enhancement effect for the detection of the diabetic foot using thermogram images. Different pre-trained networks with and without image enhancement techniques were investigated and it was found that the image enhancement techniques helped in the classification performance. The best performing Adaboost classifier can be deployed in a smartphone and can be used in the foot clinic and by users in the home setting for the early detection of DFU.

The following interesting observations can be summarized from this study:

- Gamma Correction due to its special feature enhancement has helped the network to distinguish the diabetic and control group using the dual-foot thermogram.
- A single-foot thermogram in any CNN-based classification does not improve the classification performance compared with the dual-foot approach.
- Of the various machine learning algorithms tested on the optimized feature sets the Adaboost classifier with random forest feature ranking technique outperforms all other classifiers and the 2D image-based deep learning approach.

## V. CONCLUSION

Diabetic foot ulceration has a major impact on morbidity and mortality in patients with diabetes [5]. Early detection may help to limit DFU progression and eventually amputation. The application of artificial intelligence for early detection may have considerable utility for health care professionals, especially in primary care, and for caregivers and patients to keep track of their disease. Such online solutions become more important particularly during pandemic situations where healthcare support is drastically affected due to the burden on the healthcare system. In this study, we propose a classical machine learning-based framework for the early detection of the diabetic foot from thermogram images captured using Infra-Red

(IR) cameras with a smartphone. Optimization of the thermogram features from a single foot thermogram has enabled the development of a diagnostic system that outperforms 2D image-based deep learning techniques. The proposed network can be easily deployed on a smartphone-based application and validate in a clinical trial.

### ACKNOWLEDGEMENT

This work was made possible by Qatar National Research Fund (QNRF) NPRP12S-0227-190164 and International Research Collaboration Co-Fund (IRCC) grant: IRCC-2021-001 and Universiti Kebangsaan Malaysia under Grant DPK-2021-001. The statements made herein are solely the responsibility of the authors. Open Access publication of this article is supported by Qatar National Library.